%% file: main.tex
\documentclass[12pt]{article}

\newenvironment{wileykeywords}{\textsf{Keywords:}\hspace{\stretch{1}}}{\hspace{\stretch{1}}\rule{1ex}{1ex}}

\usepackage{amsmath,amssymb}
\usepackage{graphicx}
\usepackage{color}
\usepackage{dcolumn}
\usepackage{bm}
\usepackage[numbers,super,comma,sort&compress]{natbib}
\usepackage{multirow}

\usepackage{hyperref}

\definecolor{background-color}{gray}{0.98}

\title{Coupling finite and boundary element methods to solve the Poisson--Boltzmann equation for electrostatics in molecular solvation}
\author{Micha\l{} Bosy\thanks{School of Computer Science and Mathematics, Kingston University London, Penrhyn Road, Kingston upon Thames, KT1 2EE, UK} \thanks{ORCID: 0000-0003-2723-6913}, 
    Matthew W. Scroggs\thanks{Department of Mathematics, University College London, 25 Gordon Street, WC1H 0AY London, UK} \thanks{ORCID: 0000-0002-4658-2443}, 
    Timo Betcke\footnotemark[2] \thanks{ORCID: 0000-0002-3323-2110},~\\
    Erik Burman\footnotemark[2] \thanks{ORCID: 0000-0003-4287-7241},~
    Christopher D. Cooper\thanks{Department of Mechanical Engineering and Centro Cient\'ifico Tecnol\'ogico de Valpara\'iso, Universidad T\'ecnica Federico Santa Mar\'ia, Valpara\'iso 2390123, Chile} \thanks{ORCID: 0000-0003-0282-8998}}

\begin{document}

\maketitle

\begin{abstract}
The Poisson--Boltzmann equation is widely used to model electrostatics in molecular systems. Available software packages solve it using finite difference, finite element, and boundary element methods, where the latter is attractive due to the accurate representation of the molecular surface and partial charges, and exact enforcement of the boundary conditions at infinity. However, the boundary element method is limited to linear equations and piecewise constant variations of the material properties. In this work, we present a scheme that couples finite and boundary elements for the Poisson--Boltzmann equation, where the finite element method is applied in a confined {\it solute} region, and the boundary element method in the external {\it solvent} region. As a proof-of-concept exercise, we use the simplest methods available: Johnson--N\'ed\'elec coupling with mass matrix and diagonal preconditioning, implemented using the Bempp-cl and FEniCSx libraries via their Python interfaces. We showcase our implementation by computing the polar component of the solvation free energy of a set of molecules using a constant and a Gaussian-varying permittivity. We validate our implementation against the finite difference code APBS (to 0.5\%), and show scaling from protein G B1 (955 atoms) up to immunoglobulin G (20\,148 atoms). For small problems, the coupled method was efficient, outperforming a purely boundary integral approach. For Gaussian-varying permittivities, which are beyond the applicability of boundary elements alone, we were able to run medium to large sized problems on a single workstation. Development of better preconditioning techniques and the use of distributed memory parallelism for larger systems remains an area for future work. We hope this work will serve as inspiration for future developments that consider space-varying field parameters, and mixed linear-nonlinear schemes for molecular electrostatics with implicit solvent models. 
\end{abstract}

\begin{wileykeywords}
Finite element method, Boundary element method, Poisson--Boltzmann, Implicit solvent model, Electrostatics.
\end{wileykeywords}

\clearpage


\begin{figure}[h]
\centering
\colorbox{background-color}{
\fbox{
\begin{minipage}{1.0\textwidth}
\includegraphics[width=50mm,height=50mm]{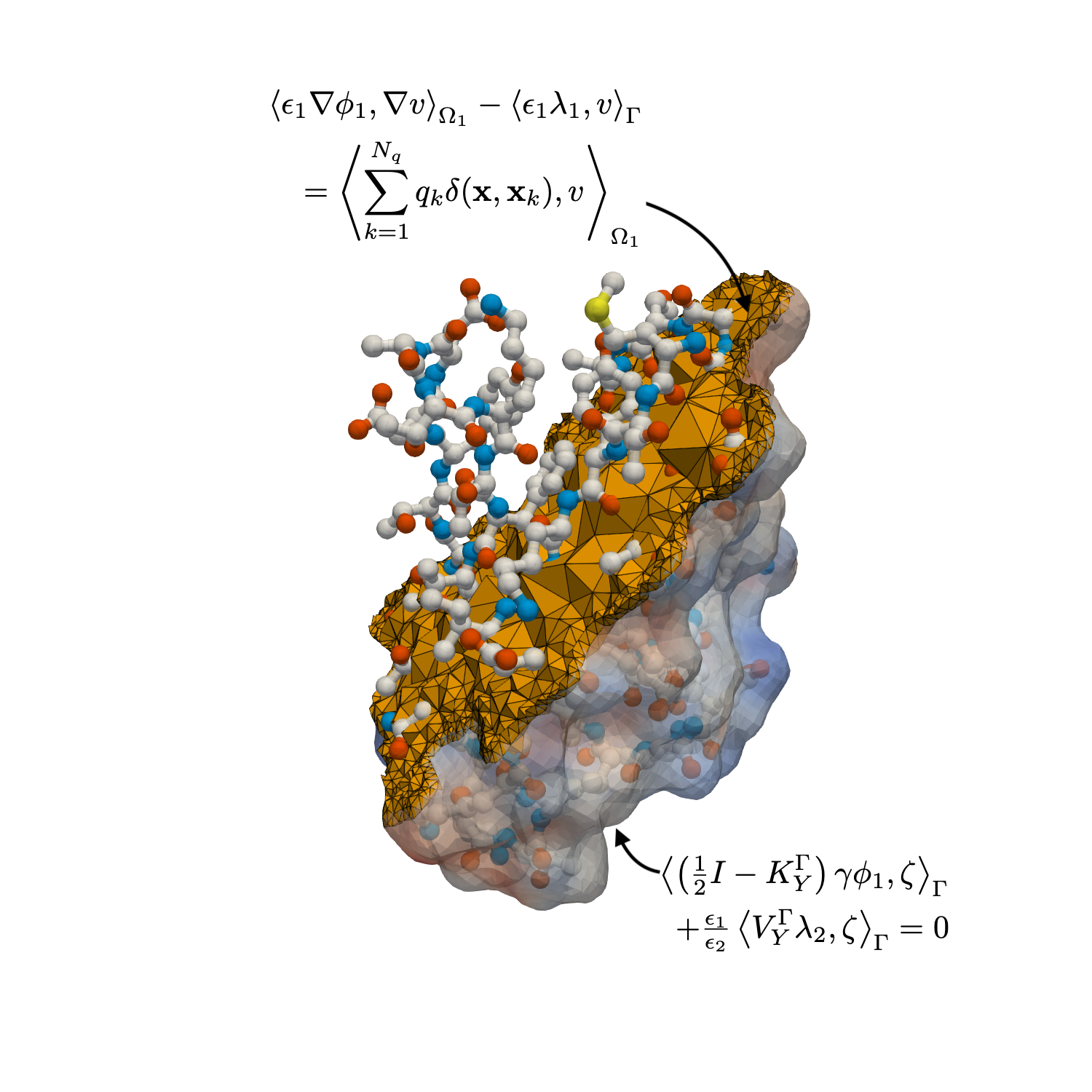} 
\\
    The boundary element method is a popular numerical algorithm to solve the Poisson--Boltzmann equation for molecular electrostatics. However, this technique is limited to linear equations and piecewise constant variations of the field parameters. Here, we overcome such limitations by coupling it with a finite element method, implemented using the Bempp-cl and FEniCSx libraries via their Python interfaces. This results in an efficient, accurate, flexible, and easy-to-use computational tool for model development. 
\end{minipage}
}}
\end{figure}

\makeatletter
  \renewcommand\@biblabel[1]{#1.}
  \makeatother

\bibliographystyle{apsrev}

\renewcommand{\baselinestretch}{1.5}
\normalsize

\clearpage

\section*{\sffamily \Large INTRODUCTION} 

\input{intro}

\section*{\sffamily \Large METHODOLOGY}
\input{methods}




\section*{\sffamily \Large RESULTS AND DISCUSSION}
\input{results}

 






\section*{\sffamily \Large CONCLUSIONS}
\input{conclusion}

\subsection*{\sffamily \large ACKNOWLEDGMENTS}
MB acknowledges the support from Kingston University through First Kingston University Grant. \\
MS acknowledges support from EPSRC grant EP/W007460/1 \\
TB acknowledges support from EPSRC grants EP/W007460/1 and EP/W026260/1. \\
EB acknowledges support from EPSRC grants EP/V050400/1 and EP/W007460/1. \\
CDC acknowledges the support from CCTVal through ANID PIA/APOYO AFB220004.\\
We would like to thank J{\o}rgen Dokken for his help with various implementation details related to FEniCSx.

\subsection*{\sffamily \large DATA AVAILABILITY STATEMENT}
All codes to reproduce the results of this manuscript can be found in the GitHub repository \url{https://github.com/MichalBosy/FEM_BEM_coupling/}, alongside links to Docker images
that include the codes alongside installations of appropriate versions of Bempp-cl and FEniCSx.


\clearpage


\bibliography{main}   

\end{document}

%% file: intro.tex
In biologically relevant settings, the structure and function of biomolecules are largely determined by the surrounding water, which usually contains salt. 
To describe these systems accurately, we need to account for the solvent correctly, which has given rise to a wide range of models.\cite{onufriev2018water}
Highly detailed models consider every water molecule and salt ion explicitly.
For solutions with large numbers of molecules, however, these models can be very compuationally expensive, so implicit-solvent models---approximated models that use continuum theory to represent the ionic solution---are often used instead.\cite{RouxSimonson1999,DecherchiETal2015}
In the case of electrostatics, the implicit-solvent model is mathematically characterized by the Poisson--Boltzmann equation (PBE)\cite{Baker2004,Bardhan2012}, which is widely used to compute solvation free energies and mean-field potentials.

The implicit-solvent model for electrostatics describes the dissolved molecule as an infinite medium with a low-dielectric solute-shaped cavity, which contains a charge distribution from the partial charges---usually a sum of Dirac deltas at the atom's locations.
The outer solvent region is represented with a high dielectric constant, and considers the presence of salt.
These two regions are interfaced by the molecular surface where the continuity of the electrostatic potential and electric displacement are enforced.
The molecular surface can be defined in various ways \cite{HarrisBoschitcshFenley2013}.

The PBE has been solved numerically with finite difference\cite{BakerETal2001,GilsonETal1985,JurrusETal2018,LiETal2019}, finite element\cite{HolstETal2012,BondEtal2010,nakov2021argos}, boundary element\cite{boschitsch2002fast,LuETal2006,AltmanBardhanWhiteTidor09,bajaj2011efficient,GengKrasny2013,CooperBardhanBarba2014}, and analytical\cite{YapHeadgordon2010,FelbergETal2017} methods.
In particular, the boundary element method (BEM) has proven to be very efficient for high-accuracy calculations \cite{GengKrasny2013,CooperBardhanBarba2014}, mainly due to the precise description of the molecular surface and point charges. 
However, BEM is limited to constant material properties in each region and the linear version of the PBE. 
Even though these limitations are acceptable in a wide range of applications, there are cases when BEM falls short: for example, if a variable permittivity is required inside the solute \cite{grant2001smooth,li2013dielectric}, or the solute is highly charged such that the linear approximation breaks\cite{FogolariETal1999}.

The present article describes a methodology to overcome some of those limitations, by coupling finite and boundary element methods.
This approach brings the best of both worlds---the flexibility of FEM and the efficiency of BEM---all in an accurate description of the solute molecule.
FEM-BEM coupling is a popular technique in the context of mixed linear-nonlinear models,\cite{carstensen1995coupling,aurada2013classical} fracture mechanics,\cite{aour2007coupled} fluid-structure interaction,\cite{estorff1991fem} acoustics,\cite{hiptmair2006stabilized} and electromagnetics.\cite{matsuoka1988calculation,hiptmair2008stabilized,bruckner20123d}
On the other hand, the PBE has been solved with hybrid numerical methods in the past: for example, by coupling finite differences with boundary elements\cite{boschitsch2004hybrid} or finite elements\cite{xie2016new,ying2018hybrid} to solve the nonlinear PBE in a specific region, and to implement modifications to the PBE model ({\it ie.} the size-modified PBE).
To the best of our knowledge, this is the first time finite and boundary elements have been combined in this application.

In this paper, we prototype this principle with the simplest implementation possible: a Johnson--N\'ed\'elec\cite{johnson1980coupling} coupling, where we solve using mass-matrix and diagonal preconditioning and whithout distributed memory execution.
This limits the size of problems we can access currently, however, it sets the basis for future developments that use more elaborate formulations and algorithms that are readily available in open source software libraries.
In this work, we use the boundary element library Bempp-cl\cite{BetckeScroggs2021} and the finite element library FEniCSx\cite{BasixJoss,BasixDofTransformations}.
These libraries are easy to use and their full functionalities may be accessed via their Python interfaces, making them ideal tools for easily implementing problems like this as well as for exploring computational efficiency and moving towards tackling large-scale problems, such as a full viral capsid.\cite{MartinezETal2019,wang2021high}
We hope this work will inspire research along these lines.

%% file: methods.tex
\subsection*{\sffamily \large The implicit solvent model}

\begin{figure}
\centering
\includegraphics[width=0.3\textwidth]{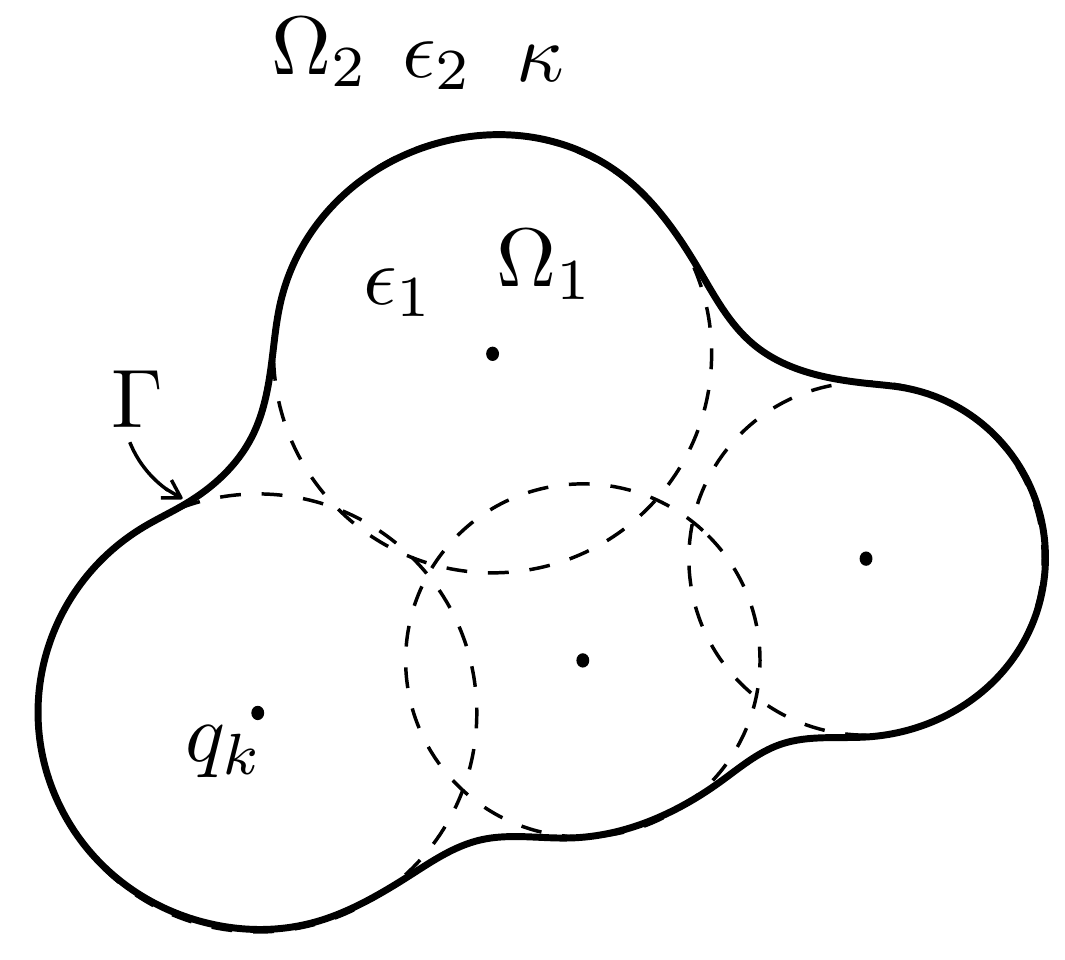}
\caption{Representation of a molecule in an implicit solvent.}
\label{fig:implicit_molecule}
\end{figure}

The implicit solvent model\cite{RouxSimonson1999,DecherchiETal2015} can be described mathematically as a coupled system of partial differential equations, where the Poisson--Boltzmann equation governs in the solvent region ($\Omega_1$ in Figure \ref{fig:implicit_molecule}), and the Poisson equation govnerns in the solute region ($\Omega_2$ in Figure \ref{fig:implicit_molecule}). These regions are interfaced by the molecular surface ($\Gamma$), where the potential ($\phi$) and electric displacement ($\epsilon\partial\phi/\partial\mathbf{n}$) are continuous across the surface.
\begin{subequations}
\label{eq:pbe}
\begin{align}
\nabla^2\phi_1(\mathbf{x}) &= \tfrac{1}{\epsilon_1}\sum_{k=1}^{N_q} q_k\delta(\mathbf{x},\mathbf{x}_k)&&\mathbf{x} \in \Omega_1,\label{eq:pbe1}\\
\left(\nabla^2 - \kappa^2\right)\phi_2(\mathbf{x})  &= 0&&\mathbf{x}\in\Omega_2,\label{eq:pbe2}\\
\phi_1(\mathbf{x})  &= \phi_2 (\mathbf{x}),&&\mathbf{x}\in \Gamma,\label{eq:pbe3}\\
\epsilon_1\frac{\partial\phi_1}{\partial\mathbf{n}}(\mathbf{x}) &= \epsilon_2\frac{\partial\phi_2}{\partial\mathbf{n}}(\mathbf{x})&&\mathbf{x}\in \Gamma. \label{eq:pbe4}
\end{align}
\end{subequations}
Here, $\epsilon_1$ and $\epsilon_2$ are the dielectric constants in the solute and solvent, respectively; $\kappa$ is the inverse of the Debye length, related to the salt concentration; and $q_k$ are the values of the partial charges, located at $\mathbf{x}_k$.

The electrostatic potential in $\Omega_1$ can be further decomposed into singular and regular components as $\phi_1 = \phi_c + \phi_r$, where $\phi_c$ is the solution to
\begin{align}\label{eq:phic}
\nabla^2\phi_c(\mathbf{x}) &= \tfrac{1}{\epsilon}\sum_{k=1}^{N_q}q_k\delta(\mathbf{x},\mathbf{x}_k) &&\mathbf{x}\in\Omega_1\cup\Omega_2,\nonumber\\
\phi_c(\mathbf{x})&=0 &&\text{ as } |\mathbf{x}|\to\infty.
\end{align}
Physically, $\phi_c$ can be interpreted as the Coulomb-type potential from the point charges, whereas $\phi_r$, also known as reaction potential, is originated by the polarization of the solvent and reorganization of the free ions. 
Usually, $\epsilon_1$ is considered a constant value, yielding an analytical expression for $\phi_c$. However, this is not the general case.

Regularized versions of the equations in \eqref{eq:pbe} \cite{LuZhouHolstMcCammon2008,LeeGengZhao2021} are widely used to numerically solve the Poisson--Boltzmann equation with finite element or finite difference methods, and have recently been extended to super-Gaussian permittivities in the solute.\cite{wang2022regularization} However, here we use the standard formulation in \eqref{eq:pbe}, as it offers more flexibility when dealing with, for example, variable permittivities, beyond super-Gaussian descriptions.

A common quantity of interest in implicit solvent models is the solvation free energy, which is the change in Gibbs free energy as the molecule moves from the vacuum into the solvent. Considering the charge distribution $\rho$ consists of point charges, this can be calculated as
\begin{equation}\label{eq:dG} 
\Delta G_{\text{solv}} = \tfrac{1}{2}\int_{\Omega_1} \rho(\mathbf{x})\phi_{r}(\mathbf{x}) = \tfrac{1}{2}\sum_{k=1}^{N_q} q_k\phi_r(\mathbf{x_k}).
\end{equation}

\subsection*{\sffamily \large BEM-BEM coupling}

Several possible boundary integral formulations of \eqref{eq:pbe} exist.~\cite{search2022towards} The simplest form was presented by Yoon and Lenhoff in 1991\cite{YoonLenhoff1990} and is known as the \emph{direct formulation}. However, the resulting system is usually ill-conditioned. Here, we use the better-conditioned formulation presented by Lu and co-workers,~\cite{LuETal2006,LuETal2009,debuhr2016dashmm} which reads
\begin{align}\label{eq:lu}
    \tfrac{\phi_2}{2}\left(1+\tfrac{\epsilon_1}{\epsilon_2}\right) - \left(K_Y^\Gamma - \tfrac{\epsilon_1}{\epsilon_2}K_L^\Gamma\right)\phi_2 + \left(V_Y^\Gamma - V_L^\Gamma\right)\lambda_2 &= \sum_{k=0}^{N_q}  \frac{q_k}{4\pi\epsilon_2|\mathbf{x}_{\Gamma} - \mathbf{x}_k|},
     \nonumber \\
    \tfrac{\epsilon_1}{\epsilon_2}\left(W_Y^\Gamma - W_L^\Gamma\right)\phi_2 +  \tfrac{\lambda_2}{2}\left(1+\tfrac{\epsilon_1}{\epsilon_2}\right) + \left(\tfrac{\epsilon_1}{\epsilon_2}K_Y^{\prime\Gamma} - K_L^{\prime\Gamma}\right)\lambda_2 &= \sum_{k=0}^{N_q}  \frac{\partial}{\partial\mathbf{n}_\mathbf{x}}\left(\frac{q_k}{4\pi\epsilon_2|\mathbf{x}_{\Gamma} - \mathbf{x}_k|}\right),
\end{align}
where $\phi_2 = \phi_2(\mathbf{x}_\Gamma)$ is the potential on $\Gamma$ as we approach from $\Omega_2$ ({\it ie.} the exterior Dirichlet trace),
and $\lambda_2 = \frac{\partial}{\partial \mathbf{n}}\phi_2$ is the normal derivative $\Gamma$ as we approach from $\Omega_2$ ({\it ie.} the exterior Neumann trace).
Note that we can obtain $\phi_1$ and $\lambda_1$ from $\phi_2$ and $\lambda_2$ by applying the interface conditions \eqref{eq:pbe3} and \eqref{eq:pbe4}.
The operators $V$, $K$, $K^{\prime}$, and $W$ that appear in Equation \eqref{eq:lu} are the single-layer, double-layer, adjoint double-layer, and hypersingular operators for the Laplace (subscript $L$) and Yukawa (subscript $Y$, also known as modified Helmholtz) kernels. These are defined 
by
\begin{align}\label{eq:all_op}
V_i^\Gamma \varphi (\mathbf{x}) &= \oint_\Gamma g_i(\mathbf{x},\mathbf{x}')\varphi(\mathbf{x}')d\mathbf{x}',\nonumber\\
K_i^\Gamma \varphi (\mathbf{x}) &= \oint_\Gamma \frac{\partial g_i}{\partial\mathbf{n}'}(\mathbf{x},\mathbf{x}')\varphi(\mathbf{x}')d\mathbf{x}',\nonumber\\
K^{\prime\Gamma}_i\varphi (\mathbf{x}) &= \oint_\Gamma \frac{g_i}{\partial\mathbf{n}}(\mathbf{x},\mathbf{x}')\varphi(\mathbf{x}')d\mathbf{x}',\nonumber\\
W^\Gamma_i\varphi (\mathbf{x}) &= - \oint_\Gamma \frac{\partial^2 g_i}{\partial\mathbf{n}'\partial\mathbf{n}}(\mathbf{x},\mathbf{x}')\varphi(\mathbf{x}')d\mathbf{x}',
\end{align}
where $i \in \{L,Y\}$, $\varphi(\mathbf{x})$ can be any distribution over $\Gamma$, and 
\begin{align}\label{eq:green_func}
g_L(\mathbf{x},\mathbf{x}')=\frac{1}{4\pi|\mathbf{x}-\mathbf{x}'|} \nonumber \\
g_Y(\mathbf{x},\mathbf{x}')=\frac{e^{-\kappa|\mathbf{x}-\mathbf{x}'|}}{4\pi|\mathbf{x}-\mathbf{x}'|}
\end{align}
are the corresponding free-space Green's functions.

Having computed the electrostatic potential with Eq.~\eqref{eq:lu}, we obtain the reaction potential ($\phi_r$) in Eq.~\eqref{eq:dG} by subtracting the Coulombic component from $\phi_1$. This gives\cite{CooperBardhanBarba2014}
\begin{align} \label{eq:phi_reac}
\phi_{r}(\mathbf{r}) = - K_{L}^{\Gamma} \phi_1(\mathbf{r}) + V_{L}^{\Gamma}  \lambda_1(\mathbf{r}),
\end{align}
where $\mathbf{r}\in\Omega_1$.
This is valid as long as $\epsilon_1$ is a constant, so we use Eq.~\eqref{eq:phi_reac} for both BEM-BEM and FEM-BEM coupling simulations with constant $\epsilon_1$.

\subsection*{\sffamily \large Novel numerical solution of the Poisson--Boltzmann equation}


The boundary element method (BEM) is a standard tool for the numerical solution of the Poisson--Boltzmann equation in molecular electrostatics.~\cite{ZauharMorgan1985, Shaw1985} This was implemented in numerous codes, such as AFMPB,~\cite{LuETal2006} TABI,~\cite{GengKrasny2013} PyGBe,~\cite{CooperBardhanBarba2014,cooper2016pygbe} and more recently, with Bempp-cl\cite{search2022towards}. One strong advantage of BEM is that it reduces the dimension of the problem by using the boundary integral formulation. Unfortunately, this advantage comes at a cost: it requires a fundamental solution to be known in order for the method to be applied. There are many linear problems for which fundamental solutions are not known: this is the case for the Poisson-Boltzmann equation with a heterogenous permittivity inside the molecule, \emph{ie}.
\begin{align} \label{eq:pbe_vp}
\nabla \cdot \left(\epsilon_1(\mathbf{x}) \nabla \phi_1(\mathbf{x})\right) &= \sum_{k=1}^{N_q} q_k\delta(\mathbf{x},\mathbf{x}_k) &&\mathbf{x} \in \Omega_1,\nonumber\\
\left(\nabla^2 - \kappa^2\right)\phi_2(\mathbf{x})  &= 0 &&\mathbf{x}\in\Omega_2,\nonumber\\
\phi_1(\mathbf{x})  &= \phi_2(\mathbf{x}),&&\mathbf{x}\in \Gamma,\nonumber\\
\epsilon_1(\mathbf{x})\frac{\partial\phi_1}{\partial\mathbf{n}}(\mathbf{x})  &= \epsilon_2\frac{\partial\phi_2}{\partial\mathbf{n}}(\mathbf{x}) &&\mathbf{x}\in \Gamma. 
\end{align}
A finite element method (FEM) is more suitable for such a case.

The coupling of FEM and BEM is an approach that can be used to solve a wide range of multiphysics problems on unbounded domains\cite{hiptmair2002,banjai2015}, as it takes advantage of both methods. On the one hand, the BEM satisfies the infinite boundary conditions exactly when they decay to zero, and only approximates boundary conditions on surfaces; hence, it is commonly used for problems involving infinite or semi-infinite domains. On the other hand, the FEM is known for its robustness and universal applicability, even for problems of inhomogeneous or non-linear nature.
Here, we use the Johnson--N\'ed\'elec formulation,\cite{johnson1980coupling} which is the simplest formulation of FEM-BEM coupling, and we detail it next.

We start with the variational formulation of the internal problem. Applying integration by parts to the first equation of~\eqref{eq:pbe_vp} we have, for every $v \in H_0^1(\Omega_1)$,
\begin{equation}
\label{eq:fem}
 \left\langle \epsilon_1 \nabla \phi_1, \nabla v \right\rangle_{\Omega_1}  -  \left\langle  \epsilon_1\partial_n \phi_1, v \right\rangle_\Gamma =  \left\langle \sum_{k=1}^{N_q} q_k\delta(\mathbf{x},\mathbf{x}_k),  v\right\rangle_{\Omega_1},
\end{equation}
where $v$ is a test function, and $\left\langle\varphi,v\right\rangle_\Gamma = \int_\Gamma \varphi(\mathbf{x})v(\mathbf{x})\,\mathrm{d}\mathbf{x}$ and $\left\langle\varphi,v\right\rangle_{\Omega_1} = \int_{\Omega_1} \varphi(\mathbf{x})v(\mathbf{x})\,\mathrm{d}\mathbf{x}$ are the inner products on the surface and in the domain, respectively. Slightly abusing notation, the product $\left\langle \cdot, \cdot \right\rangle_{\Omega_1}$ on the right-hand side denotes the duality paring.
For the external problem with BEM, we define the Dirichlet trace~\cite{MR2361676} 
\begin{align*}
\gamma:  H^1(\Omega_2) &\rightarrow H^{\frac{1}{2}}(\Gamma), & \gamma f(x) & := \lim_{\Omega_2 \ni \mathbf{y} \rightarrow \mathbf{x} \in \Gamma}  f(\mathbf{y}),
\end{align*}
and we use the direct formulation from the second equation of Eq.~\eqref{eq:pbe} to obtain
\begin{align*}
\tfrac{\phi_1}{2} - K_{Y}^{\Gamma}\gamma \phi_1 + \tfrac{\epsilon_1}{\epsilon_2}V_{Y}^{\Gamma}  \lambda_1 & = 0.
\end{align*}
$V$ and $K$ are the single-layer and double-layer operators as defined in Equation~\eqref{eq:all_op}.
%
%
%
%

Then, the coupling problem can be written as: \textit{Find $ \phi_1 \in H^1(\Omega_1)$ and $\lambda_1 \in H^{-\frac{1}{2}}(\Gamma)$ such that for all $v \in H^1(\Omega_1)$ and $\zeta \in H^{-\frac{1}{2}}(\Gamma)$},
\begin{subequations}
\label{eq:standard_fem_bem}
\begin{align} 
 \left\langle  \epsilon_1 \nabla \phi_1, \nabla v \right\rangle_{\Omega_1}  - \left\langle \epsilon_1 \lambda_1, v \right\rangle_\Gamma &=   \left\langle  \sum_{k=1}^{N_q} q_k\delta(\mathbf{x},\mathbf{x}_k),  v \right\rangle_{\Omega_1}, \\[3mm] 
  \left\langle \left\langle\tfrac{1}{2} I - K_{Y}^{\Gamma}\right) \gamma \phi_1, \zeta \right\rangle_\Gamma + \tfrac{1}{\epsilon_2} \left\langle V_{Y}^{\Gamma}\epsilon_1\lambda_1, \zeta \right\rangle_\Gamma &=0.
\end{align}
\end{subequations}
Note that in the case where $\epsilon_1$ is constant, then each $\epsilon_1$ can be moved outside of the inner product it is inside.


When discretised, this can also be written in matrix form. Let $\vec{\phi}_1 := [\phi_1^1, \dots, \phi_1^j]^T$ be the vector of canonical basis
functions of the finite element space $V_{h}^{j}\subset H^1(\Omega_1)$, 
and let $\vec{\lambda}_1 := [\lambda_1^1, \dots, \lambda_1^l]^T$ be the vector of canonical basis
functions of $\Lambda_{h}^{l}\subset H^{-\frac{1}{2}}(\Gamma)$. 
We define the following matrices associated with the corresponding bilinear forms
\begin{align*}
A_{\alpha \beta} &= \left\langle\epsilon_1\nabla\phi_1^{\alpha}, \nabla \phi_1^{\beta} \right\rangle_{\Omega_1},&
\widetilde{M}_{\alpha \beta} &= \left\langle \epsilon_1\lambda^{\beta}_1, \gamma \phi_1^{\alpha}\right\rangle_{\Gamma},\\
K_{\alpha \beta} &= \left\langle K_{Y}^{\Gamma} \gamma \phi^{\alpha}_1, \lambda^{\beta}_1 \right\rangle_{\Gamma},&
V_{\alpha \beta} &= \left\langle V_{Y}^{\Gamma} \epsilon_1\lambda^{\alpha}_1, \lambda^{\beta}_1 \right\rangle_{\Gamma},\\
M_{\alpha \beta} &= \left\langle \gamma \phi^{\alpha}_1, \lambda^{\beta}_1 \right\rangle_{\Gamma},
\end{align*}
and vector associated with the corresponding linear form
\begin{equation*}
\vec{f}_{\beta} := \left\langle  \sum_{k=1}^{N_q} q_k\delta(\mathbf{x},\mathbf{x}_k),  \phi_1^{\beta} \right\rangle_{\Omega_1}.
\end{equation*}
Using the above definitions, the discrete problem in \eqref{eq:standard_fem_bem} can be written in the following blocked matrix form:
\begin{align}\label{eq:fembem_matrix}
\begin{bmatrix}
A & -\widetilde{M}^T \\  
\tfrac12 M - K &  \tfrac{1}{\epsilon_2} V 
\end{bmatrix}
\begin{bmatrix}
\vec{\phi}_1 \\  
\vec{\lambda}_1
\end{bmatrix}
= 
\begin{bmatrix}
\vec{f} \\  
0
\end{bmatrix}.
\end{align}

%% file: results.tex
This section presents the verification and performance results of the presented FEM-BEM coupling schemes for molecules modeled as cavities with constant and varying permittivity.
With a constant permittivity inside the molecule, we tested convergence against an analytical expression of the solvation energy of a sphere \cite{Kirkwood1934}, and then compared a more realistic geometry (arginine) with a purely BEM implementation.
We also considered a Gaussian-varying permittivity\cite{grant2001smooth,li2013dielectric} inside the molecular cavity of arginine, and used APBS \cite{BakerETal2001} to verify our results.
The final tests show the scaling of the FEM-BEM coupling as the molecule size grows. 

All runs were done on a Lenovo ThinkStation P620 with AMD Ryzen ThreadRipper PRO 3975WX (32-core and 3.5 GHz) and 128 GB RAM. 

\section*{\sffamily \Large Software environment}

For the finite element computations, we use the software package FEniCSx~\cite{BasixJoss,BasixDofTransformations} while for the boundary element computations, we use Bempp-cl~\cite{BetckeScroggs2021} together with Exafmm-t~\cite{exafmm-t}. FEniCSx is the successor of the widely used FEniCS finite element library\cite{FEniCS,FEniCS-book}.
It provides a convenient Python interface, describing problems using Unified Form Language (UFL)~\cite{UFL}, a convenient domain-specific language specifically designed for finite element discretisations of partial differential equations. During assembly, the UFL description is transformed into efficient low-level C++ code and just-in-time compiled~\cite{ffc1,ffc2}. Bempp-cl is a Python package that uses low-level OpenCL kernels written in C99 to provide optimised assembly routines~\cite{2021-cise}. The built-in dense assembly routines are highly efficient for moderate discretisation sizes up to a few ten thousand elements.

For very large grid sizes the user can enable fast multipole method (FMM) assembly which internally is handled in Bempp-cl through an interface to the Exafmm-t FMM library. For $N$ surface elements this reduces the memory and computational complexity from $\mathcal{O}(N^2)$ in the dense assembly case to $\mathcal{O}(N)$ in the FMM case, making large boundary element problems tractable on standard workstations. In this work, we only use the FMM in the performance analysis for larger structures, as all other cases are small enough to run efficiently with a dense assembler. 

To couple FEniCSx with Bempp-cl we load a volume mesh with FEniCSx. We then export the corresponding boundary mesh into Bempp-cl and assemble the boundary spaces there. Bempp-cl provides numerical trace operators that can translate from the degree of freedom (DOF) representation in FEniCSx to the DOF representation in Bempp-cl. The corresponding translation work is handled opaquely and the user only needs to deal with high-level interfaces of FEniCSx operators, Bempp-cl operators, and trace operators. FMM assembly fits automatically into this framework and can be enabled or disabled as a simple configuration option.
Once the discrete block matrices in Eq.~\eqref{eq:fembem_matrix} are built with Bempp-cl and FEniCSx, we assemble them into one matrix and solve the linear system with Scipy's\cite{2020SciPy-NMeth} GMRES, in this case, with a tolerance of $10^{-5}$.

Docker images containing FEniCSx, Bempp-cl, and Exafmm-t are publicly available (\url{https://bempp.com/installation.html}), and all codes used to generate the results in this section are available as Jupyter Notebooks that can be reproducibly executed in an appropriate environment. The results in this section were obtained using version 0.6.0 of FEniCSx and version 0.3.0 of Bempp-cl. All codes to reproduce the results of this manuscript can be found in the GitHub repository \url{https://github.com/MichalBosy/FEM_BEM_coupling/}.

\section*{\sffamily \Large Results with constant permitivitty}

In implicit-solvent models, the molecule is usually considered as a region with constant permittivity: in our computations, we use $\epsilon_1=2$.
In the solvent region, we used the permittivity of water ($\epsilon_2=80$) and an inverse of the Debye length of $\kappa=0.125\,\text{\AA}^{-1}$.
In this case, there is a known analytical solution for $\phi_c$ in Eq.~\eqref{eq:phic}, so it is enough to compute $\phi_r$. We do this using Eq.~\eqref{eq:phi_reac} with both BEM-BEM and FEM-BEM coupling approaches. For FEM-BEM, the integral over $\Gamma$ in Eq.~\eqref{eq:phi_reac} corresponds to the trace of the solution vector from Eq.~\eqref{eq:fembem_matrix}.

\subsection*{\sffamily \large Convergence of a spherical cavity}

\begin{figure}
  \centering
  \includegraphics[width=0.45\linewidth]{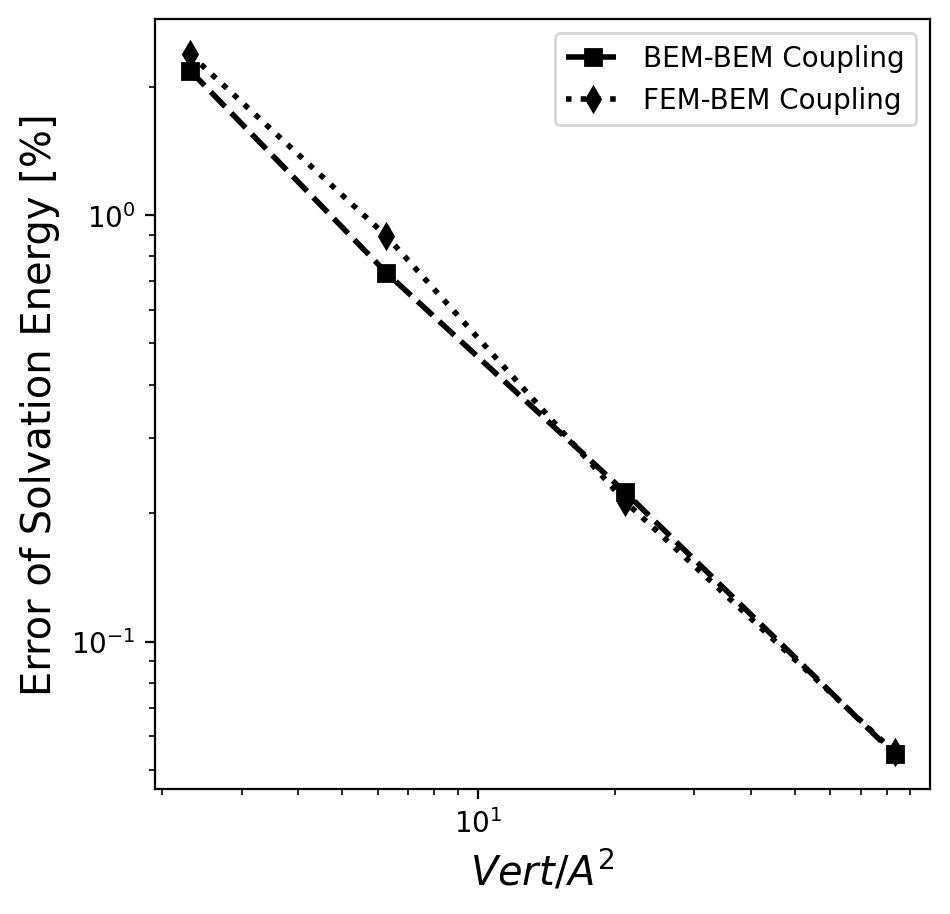}
  \caption{Error for the Kirkwood sphere.  }
  \label{fig:error_sphere}
\end{figure}

The Kirkwood sphere \cite{Kirkwood1934} is a standard benchmark test for the Poisson--Boltzmann equation in molecular electrostatics. 
In this case, we considered a spherical cavity of radius $R=2\,\text{\AA}$, with three charges ($q_1=1$, $q_2=1$, and $q_3=0.75$) placed at $\mathbf{x}_1=(1,0,0)$, $\mathbf{x}_2=(0.7,0.7,0)$, and $\mathbf{x}_3=(-0.5,-0.5,0)$.
Figure \ref{fig:error_sphere} shows the percentage error of the FEM-BEM approach, and a reference BEM-BEM implementation, compared to the analytical solution ($\Delta G_{\text{solv}}= -336.0396\,\text{kcal/mol}$).
In these runs, the FEM mesh was generated using GMSH~\cite{geuzaine2009gmsh} 
with 
2, 6, 21 and 83 vertices per $\text{\AA}^2$ on the SES. The triangular mesh on the boundary surface of the FEM mesh was used for the BEM runs. 
The error in Figure \ref{fig:error_sphere} decays linearly with the number of vertices on the surface. This agrees with the expected convergence for P1 elements, indicating a correct implementation of the numerical scheme. 

\subsection*{\sffamily \large Performance with arginine}

As a more realistic test, we assessed the performance of the FEM-BEM coupling technique against a BEM-BEM implementation for arginine.
The structure of arginine was taken from the protein data bank, and parameterized with the Amber\cite{ponder2003force} force field. 
We generated surface meshes containing 4.1, 6.7, 8.6, 17, and 24.5 vertices per $\text{\AA}^2$ with Nanoshaper.~\cite{decherchi2013general}
These densities correspond to a grid-scale parameter in Nanoshaper equal to 1.6, 2.0, 2.4, 3.4, and 4.0, respectively, where the grid scale is the reciprocal of the average characteristic length of the triangles.
For our BEM-BEM solver, we used these meshes directly. For the FEM-BEM solver, we created volume meshes from these using pyGAMer,~\cite{lee2020open} which invoked TetGen\cite{hang2015tetgen} with a quality parameter (radius-edge ratio) of 1.0.

\begin{figure}
\centering
   \includegraphics[width=0.45\linewidth]{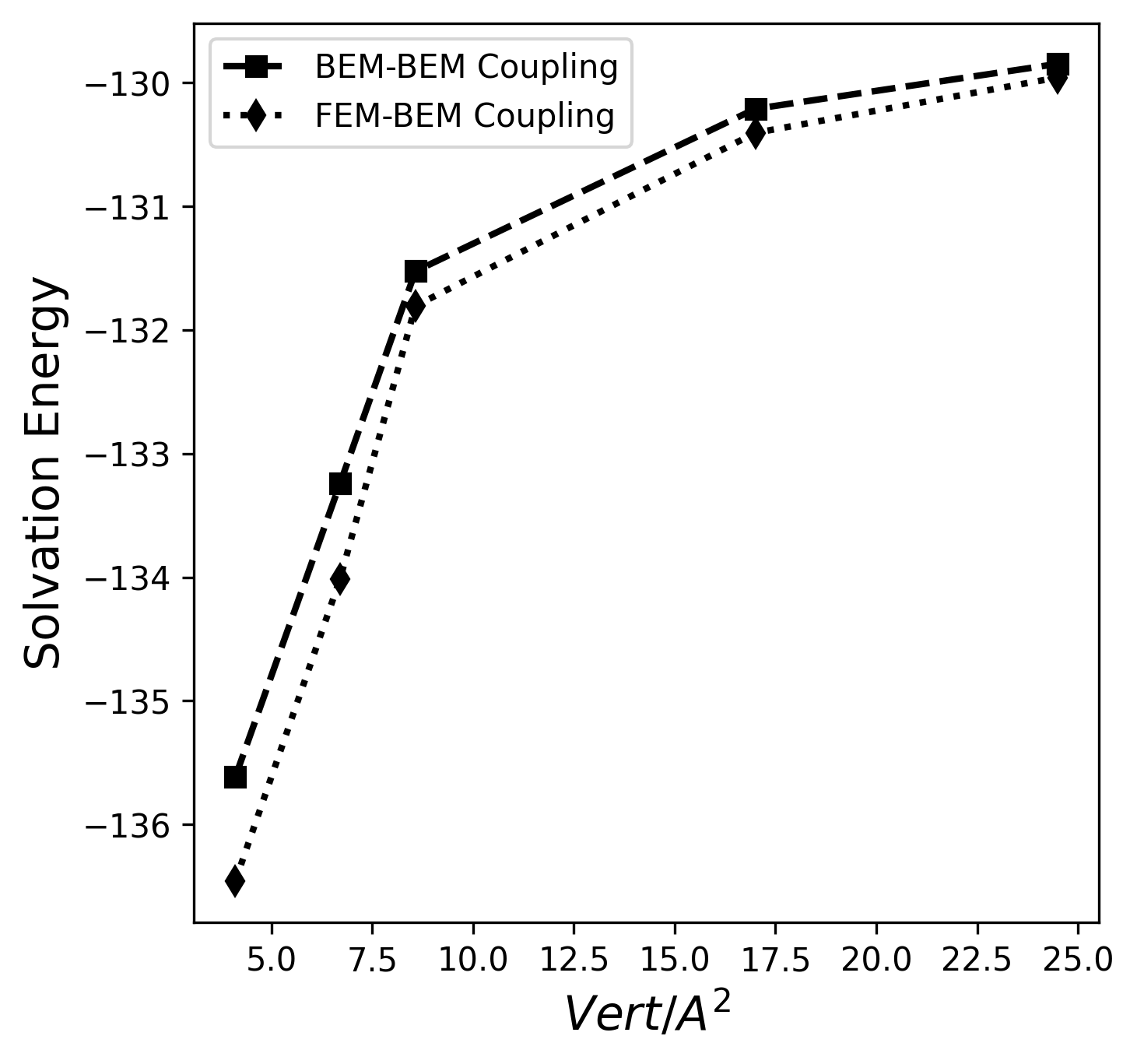}
\caption{Solvation energy for arginine with a constant permittivity.}
\label{fig:arg_constant_energy}
\end{figure}

\begin{figure}
\centering
   \includegraphics[width=0.45\linewidth]{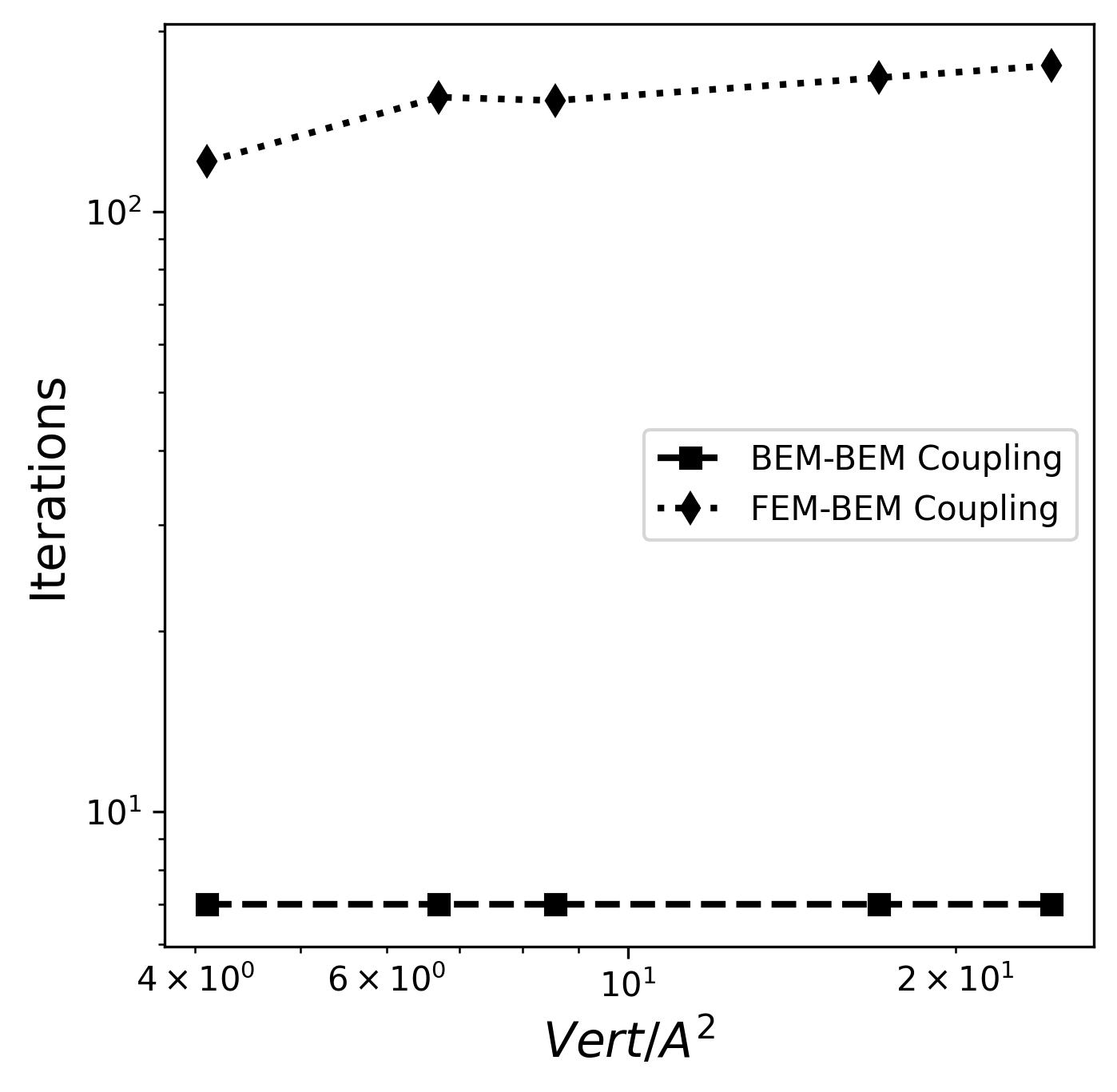}
  \includegraphics[width=0.45\linewidth]{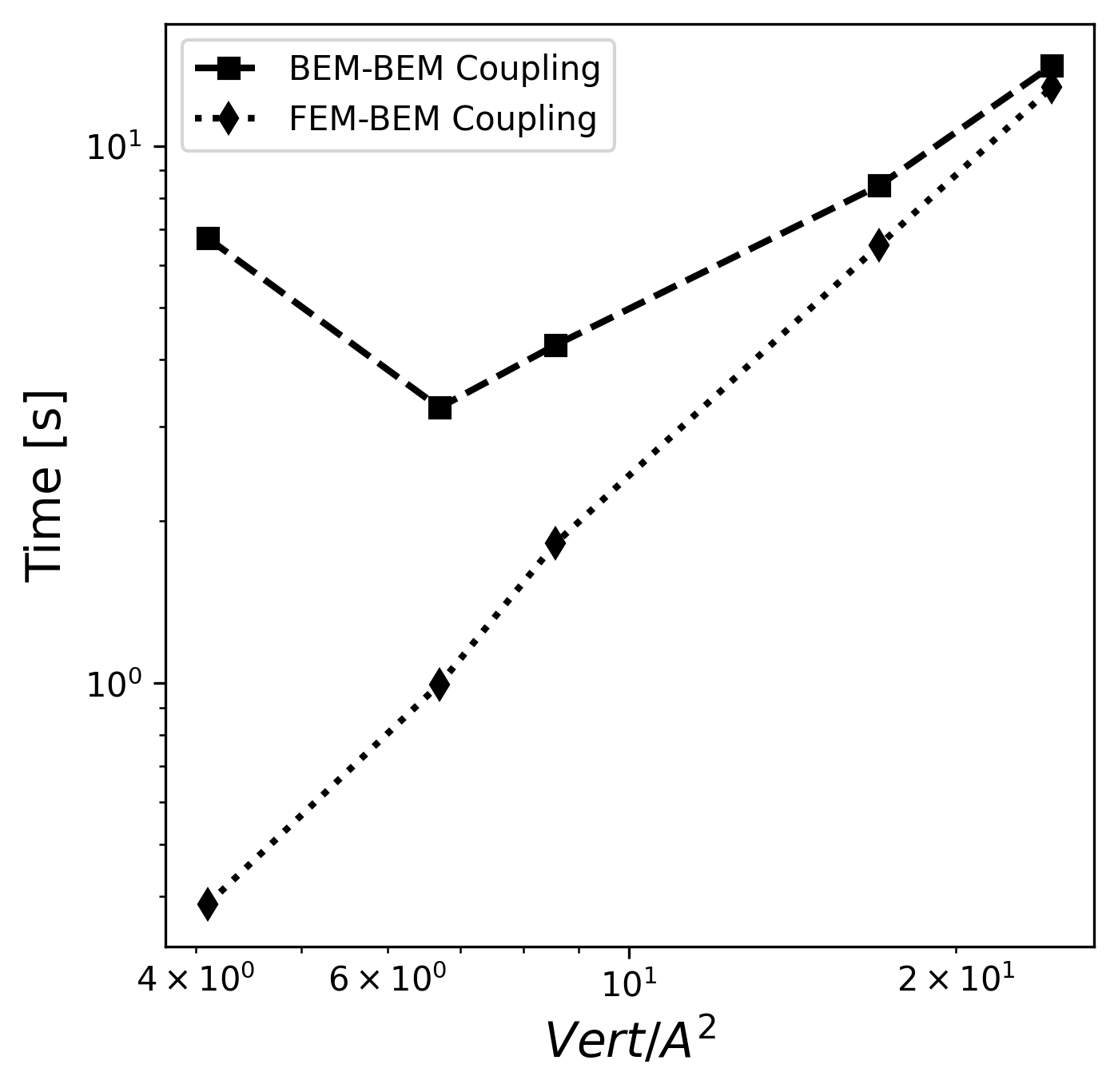}
  \caption{Iteration count (left) and time-to-solution (right) for arginine with a constant permittivity.  }
\label{fig:arg2_constant_time_iter}
\end{figure}

The solvation energy computed with the two schemes is presented in Figure \ref{fig:arg_constant_energy}, which, as expected, converges to a similar answer as the mesh is refined.
Figure \ref{fig:arg2_constant_time_iter} compares the iteration count and time-to-solution. The left plot shows that using only BEM outperforms the coupled FEM-BEM approach in terms of the iteration count. However, if we look at the total time that solvers take to obtain the solution, we can see the advantage of using the FEM-BEM coupling. The higher computational cost is caused by the need to use a hypersingular operator in the BEM formulation, and the fact that we are not using any acceleration method ($ie.$ FMM).
The timings for the FEM-BEM coupling scale are at a greater rate than the pure BEM counterpart, indicating that work on preconditioners and other acceleration methods will be required in order to make
the FEM-BEM approach viable for large problems.

\section*{\sffamily \Large Results with variable permittivity}

\subsection*{\sffamily \large Motivation: modeling the solute with a Gaussian-based variable permittivity}

In contrast to a purely BEM approach, FEM-BEM coupling gives the flexibility to consider space-varying field parameters. 
A popular description of the molecule is to consider a permittivity that varies like a Gaussian around each atom,\cite{grant2001smooth} which has shown enhanced accuracy in some applications, like pKa calculations.\cite{li2013dielectric}
In this setting, we define a density function $\rho$ depending on the position $\mathbf{r}$ as
\begin{equation}
\rho(r) := \prod_i \left(1 - \exp{\left(\frac{\|\mathbf{r}-\mathbf{x}_i\|}{\sigma^2 R_i^2}\right)}\right),
\end{equation}
where the product runs over all the atoms of the solute, $R_i$ is the van der Waals radius of atom $i$, and we used $\sigma=1$. Then, we can compute the permittivity as
\begin{equation}\label{eq:varying_eps}
\epsilon := \left(1-\rho \right) \epsilon_1 + \rho\epsilon_2
\end{equation}

As $\epsilon$ is variable, Equation \eqref{eq:phic} does not have a known analytical solution, and the electrostatic potential in the vacuum state has to be computed numerically.
For vacuum calculations, we considered the same distribution of $\epsilon$ inside the molecule as in the solvated case, but the solvent permittivity was set to $\epsilon_2=2$. 
Other implementations of Gaussian permittivities also modify the solute permittivity in vacuum calculations, according to a set cutoff.\cite{li2013dielectric} We did not consider a cutoff in our calculations.

\subsection*{\sffamily \large Convergence for arginine with APBS}

We used Equation \eqref{eq:varying_eps} to generate dielectric maps, which we ran on APBS~\cite{BakerETal2001} for comparison. 
We chose APBS because it provides an easy interface to control dielectric maps to ensure their agreement with the maps imposed in our FEM-BEM coupled approach.

Table \ref{table:arg_variable} shows a comparison of the solvation energy computed with APBS and our FEM-BEM coupling approach. We can see that they are both converging to equivalent values, where the solutions on the finest meshes agree to within 0.5\% (0.5 kcal/mol). The coarsest meshes used in both cases are within the recommended densities for accurate solvation energy simulations with constant permittivity: a finite difference mesh with $h=0.5\,\text{\AA}$ or less is recommended for binding energy calculations\cite{sorensen2015comprehensive} (the result of subtracting two solvation energies), and a similar study with BEM\cite{CooperBardhanBarba2014} showed that a mesh with 2 vertices per $\text{\AA}^2$ gives acceptable results when computing binding and solvation energies. We see a jump in the solvation energy for the FEM-BEM results going from 4.1 to 6.7 vertices per $\text{\AA}^2$, but later the energy monotonically decreases, converging to a solution. This is an indication that mesh requirements for variable permittivities may be tighter than with a constant permittivity, even though the jump in dielectric constant across the molecular surface is usually smaller. 

\begin{table}
\centering
\footnotesize
\begin{tabular}{c|c|c}
Mesh size (\AA)  & Grid points & \parbox{2cm}{\centering$\Delta G_{\text{solv}}$\\(kcal/mol)}\\
\hline
$0.52\times0.52\times0.52$ & $97\times97\times97$ & $-107.6186$\\ 
$0.39\times0.39\times0.39$ & $129\times129\times129$ & $-107.8752$\\ 
$0.26\times0.26\times0.26$ & $193\times193\times193$ & $-108.3378$\\ 
$0.195\times0.195\times0.195$ & $257\times257\times257$ & $-108.5837$\\ 
$0.098\times0.098\times0.098$ & $513\times513\times513$ & $-108.8844$\\ 
\end{tabular}%
\hspace{3mm}%
\begin{tabular}{c|c|c}
\parbox{2.1cm}{\centering Mesh density\\(vert/$\text{\AA}^2$)}& DOFs & \parbox{2cm}{\centering$\Delta G_{\text{solv}}$\\(kcal/mol)}\\
\hline
$4.1$ & $3\,491$ & $-109.931$ \\
$6.7$ & $5\,787$ & $-110.237$ \\
$8.6$ & $8\,844$ & $-109.661$ \\
$17.0$ & $19\,911$ & $-109.369$ \\
$24.5$ & $32\,302$ & $-109.315$ \\
\end{tabular}
\caption{Solvation energy of arginine with a Gaussian-like permittivity, computed using APBS (left) and the FEM-BEM approach (right). The mesh density for FEM-BEM corresponds to the vertex density of the surface mesh used to generate the volumetric mesh.}
\label{table:arg_variable}
\end{table}

\subsection*{\sffamily \large Performance analysis for larger structures}

In the experiments presented so far, we have only tested the FEM-BEM coupled approach with small structures. In this section, we study its behaviour with larger structures to evaluate its applicability in more realistic settings, and in test cases where BEM-BEM coupling is not an alternative. Using the same Gaussian permittivity field inside the molecule, we computed the solvation free energy of protein G B1 (955 atoms, PDB code 1pgb), lysozyme (1960 atoms, PDB code 1lyz), the barnase-barstar complex (9464 atoms, PDB code 1x1u), and immunoglobulin G (20148 atoms, PDB code 1igt). All structures were parameterized with the Amber\cite{Swanson05} force field using pdb2pqr.\cite{Dolinsky04} The surfaces were meshed with Nanoshaper\cite{decherchi2013general} using a grid scale of 1.5, and then volume meshes were generated using pyGAMer\cite{lee2020open} and TetGen\cite{hang2015tetgen} with a radius-edge ratio of 1.0. The solvation free energy and timings for these runs are presented in Table \ref{table:large_variable}. These results demonstrate that our FEM-BEM coupling implementation can reach medium-to-large-sized proteins on a simple workstation. For these larger structures, we enabled the FMM capabilities of Bempp-cl.

Table \ref{table:large_variable} shows that the iteration count increases with the problem size. This is expected for Johnson--N\'ed\'elec coupling, which is the simplest coupling strategy. To isolate the effect of the increased iterations in the analysis, we separate timings into the setup and solving time, where the setup time is independent of the number of iterations, and the solving time corresponds to the time spent in the GMRES solver. The time-per-iteration is computed by dividing the solving time by the number of iterations.
As we increase the number of degrees of freedom (DOFs), the setup time scales slightly worse than $\mathcal{O}(N^2)$ (where $N$ is the number of DOFs) but represents the smaller fraction of the total time.
On the other hand, the time per iteration scales closer to $\mathcal{O}(N)$, which is expected as we are using FMM for the BEM portion of the matrix.
This is an indication that the high solving time is mainly due to the increase in iteration count, and having better-conditioned coupling methods, such as the so-called hybrid approach\cite{betcke2022hybrid}, or more effective preconditioners for the blocked system would have a large impact on the time to solution.

Even though this scheme is capable of calculating the electrostatic potential in medium-to-large proteins, the largest test case in Table \ref{table:large_variable} (1igt) used up 30\% of the available RAM memory. 
If we were aiming at larger structures, such as full viruses,\cite{MartinezETal2019,wang2021high} we would require the use of optimized fast algorithms\cite{wang2021exafmm,kailasa2023pyexafmm} and
parellelization of the storage and solver, alongside the necessary improvements in the conditioning of the system.

\begin{table}
\centering
\footnotesize
\begin{tabular}{c|c|c|c|c|c|c|c|c}
Molecule & \parbox{1.1cm}{\centering FEM DOFs}  & \parbox{1.1cm}{\centering BEM DOFs} &  \parbox{1.8cm}{\centering $\Delta G_{\text{solv}}$ (kcal/mol)}& Iterations & \parbox{1.3cm}{\centering Setup time (s)} & \parbox{1.3cm}{\centering Solving time (s)} & \parbox{1.5cm}{\centering Time per iter. (s)} & \parbox{1.4cm}{\centering Total time (s)} \\[3mm]
\hline
  1pgb  & $29\,434$ &  $10\,058$ &  $-300.888$ & $703$ & $86$ & $472$ & $0.67$ & $789$  \\
 1lyz  & $56\,114$ & $18\,606$ &  $-599.310$ & $1\,073$ & $336$ & $1\,350$  & $1.26$ & $1\,686$   \\
 1x1u & $263\,181$ & $81\,258$ & $-1\,982.085$ & $2\,791$ & $8\,340$ & $11\,000$ & $3.94$ & $19\,340$  \\
 1igt & $597\,575$ & $187\,712$ &  $-3\,294.0157$ & $5\, 366$ & $41\,998$ & $40\,100$ &  $7.47$ & $82\,098$    \\
\end{tabular}
\caption{Results for larger structures with a variable permittivity.}
\label{table:large_variable}
\end{table}

%% file: conclusion.tex
This paper presents the first implementation of a FEM-BEM coupling approach to solve the Poisson--Boltzmann equation for molecular electrostatics. This brings the best of both worlds: the accuracy and efficiency of BEM to exactly enforce the boundary conditions at infinity, and the flexibility of FEM to account for space variations of the material properties and nonlinearities. After presenting verification results for a sphere and arginine with a constant permittivity inside the solute, we showcased our implementation with an advanced modeling technique that considers Gaussian-varying permittivities in a confined region, with the results validated against the widely-used APBS software. The final scaling results for larger molecules start from protein G B1 (955 atoms) and go up to immunoglobulin G (20\,148 atoms), proving the applicability of this approach for realistic problems. Even though our implementation was able to reach medium-to-large systems, we recongnize the need for further research towards better preconditioning of the linear system, optimizing the coupling technique, acceleration algorithms, and parallel execution, especially as we look towards much larger solutes, like viruses. We hope this proof-of-concept work will serve as motivation for future model development that considers space-varying permittivities and Debye lengths, and mixed linear-nonlinear techniques, especially for highly-charged systems, like nucleic acids. 